\documentclass[twocolumn,oneside]{article}

\usepackage[utf8]{inputenc}
\usepackage{epsfig}
\usepackage{booktabs}
\usepackage{natbib}
\usepackage{bm}

\def\ve#1{{\bm #1}}
\def\matrix#1{{\bf #1}}
\def\muas{\hbox{$\mu$as}}

\begin{document}

\title{Consistent modeling of the geodetic precession in Earth rotation}
\author{E.~Gerlach, S.~Klioner and M.~Soffel\thanks{Lohrmann Observatory, Dresden Technical University, 01062 Dresden, Germany} }

\maketitle

\thispagestyle{empty}

\begin{abstract}
  A highly precise model for the motion of a rigid Earth is indispensable
  to reveal the effects of non-rigidity in the rotation of the 
  Earth from observations. To meet the accuracy goal of modern
  theories of Earth rotation of 1 microarcsecond (\muas) 
  it is clear, that for such a model also relativistic effects have to
  be taken into account. The largest of these effects is the so
  called geodetic precession.\\
  In this paper we will describe this effect and the standard
  procedure to deal with it in modeling Earth rotation up to now. With
  our relativistic model of Earth rotation \citep{klioner_etal2001}
  we are able to give a con\-sistent post-Newtonian treatment of the
  rotational motion of a rigid Earth in the framework of General
  Relativity. Using this model we show that the currently
  applied standard treatment of geodetic precession is not
  correct. The inconsistency of the standard treatment leads to errors
  in all modern theories of Earth rotation with a magnitude of up to
  200 \muas\ for a time span of one century.
\end{abstract}

\section{Introduction}

Geodetic precession/nutation is the largest relativistic effect in
Earth rotation. This effect has been discovered already a few years
after the formulation of General Relativity \citep{deSitter} and very
early it was recognized to be important for Earth rotation. It results
mainly in a slow rotation of a geocentric locally inertial reference
frame with respect to remote celestial objects roughly about the
ecliptic normal. Due to its relatively large magnitude of about
$1.9^{\prime\prime}$ per century, which is $3\times10^{-4}$ of the
general precession, corresponding corrections are used in all standard
theories of precession and nutation since the IAU 1980 theory
\citep{IAU1980}.

The standard way to consider geodetic precession
in Earth rotation theories up to now was the following: firstly, 
using purely Newtonian equations, one computed 
the orientation of the Earth in a geocentric, locally inertial reference frame.
To obtain the solution with respect to the kinematically
non-rotating Geo\-centric Celestial Reference System (GCRS) the corrections for geodetic
precession were then simply added, as described for example in \citet[Section 8]{SMART1}. 
These corrections 
can be calculated separately, since they are completely independent of the rotational
state of the Earth, e.~g. \citet{brumberg_etal1991}.

The purpose of this work is to demonstrate that the standard
way of applying the geodetic precession is not correct. After stating the
problem in the following Section, we explain shortly our
relativistic model of Earth rotation used for this study in Section
\ref{sec:3}. In Section \ref{sec:4} we describe
how the corrections for the geodetic precession can be computed, while
in Section \ref{sec:5} two different, but equivalent and correct ways
to obtain a GCRS solution are given. In the last section of this paper
we compare our solution to published ones and draw concluding remarks.

Throughout the paper we will use the following conventions:
\begin{itemize}
 \item[-] Lower case Latin indices take the values $1,2,3$.
 \item[-] Repeated indices imply the Einstein's summation irrespective of their positions, e.~g. $x^iy^i=x^1y^1+x^2y^2+x^3y^3$
 \item[-] $\epsilon_{abc}$ is the fully antisymmetric Levi-Civita symbol, defined as $\epsilon_{abc}=(a-b)(b-c)(c-a)/2$.
 \item[-] Vectors are set boldface and italic: $\ve{X}=X^{\,i}$, while matrices are set boldface and upright: $\matrix{P}=P^{\,ij}$
 \item[-] The choice to use index or vector notation for a specific formula is done with regard to readability and clarity.
\end{itemize}

\section{The GCRS and geodetic precession}

The Geocentric Celestial Reference System is officially adopted by the
IAU to be used to describe physical phenomena in the vicinity of the
Earth and, in particular, the rotational motion of the Earth.  The
GCRS is connected with the Barycentric Celestial Reference System
(BCRS) by a generalized version of the Lorentz transformation. This
transformation was chosen in such a way that the GCRS spatial
coordinates $\ve{X}_K$ are kinematically non-rotating with respect to
the BCRS coordinates, i.~e. no additional spatial rotation of the
coordinates is involved in the transformation from one system to the
other \citep{soffel_etal2003}.

Since the origin of the GCRS coincides with the geocenter and the
Earth is moving in the gravitational field of the Solar system 
a local inertial frame with spatial coordinates $\ve{X}_D$
slowly rotates in the GCRS:
\begin{equation}
\label{xK2xD}
X^i_D=R^{\,ij}(T)\,X^j_K.
\end{equation}
\noindent
Here $R^{\,ij}$ is an orthogonal matrix, the time variable $T$
is the Geocentric Coordinate Time ${\rm TCG}$. 
Due to this rotation the equations
of motion in the GCRS contain a Coriolis force. The locally inertial
analogon of the GCRS is called dynamically non-rotating. This rotation
between the kinematically non-rotating GCRS and its dynamically
non-rotating counterpart is called geodetic precession.  The angular
velocity of geodetic precession $\bm \Omega_{\rm GP}$ is given by
\begin{equation}
\label{Omega-GP}
 \bm \Omega_{\rm GP} \approx \frac{1}{c^2}\sum_A 
\frac{GM_A}{r_{EA}^3}\left[\left(\frac{3}{2}\bm v_E -2\bm v_A \right)\times \bm r_{EA} \right],
\end{equation}
\noindent
where $c$ is the speed of light in the vacuum, $G$ the gravitational
constant, $\bm v_A$ the BCRS velocity of the body $A$ with mass
$M_A$, $\bm v_E$ is the velocity of the geocenter, $\bm r_{EA}$ the
vector from body $A$ to the geocenter and $r_{EA}$ its Euclidean norm.
The angular velocity $\bm{\Omega}_{\rm GP}$
corresponds to the orthogonal matrix $R^{\,ij}$ so that the respective
kinematical Euler equations read
\begin{equation}
\label{Omega-GP-R}
\Omega_{\rm GP}^a={1\over2}\,\varepsilon_{abc}\,R^{\,db}(T)\,{d\over dT}\,R^{\,dc}(T).
\end{equation}
This equation can be easily verified by direct substitution of the matrix elements.

\section{\label{sec:3}Model of Earth rotation}

A complete and profound discussion of our relativistic model of Earth
rotation can be found in
\citet{klioner_etal2001,klioner_etal2010}. For the purposes of this
work we neglect all other relativistic effects except for the geodetic
precession. In particular, we neglect relativistic torques, relativistic time
scales, and relativistic scaling of various para\-meters. Then, in 
dynamically non-rotating coordinates $\ve{X}_D$ the equations of
rotational motion of the Earth can be written as
\begin{eqnarray}
&&{d\over dT}\,\ve{S}_D=\ve{L}_D, \label{eqm-D} \\
&&\ve{S}_D=\ve{S}_D(\matrix{P}_D; {\cal A,B,C}), \label{S-D} \\
&&\ve{L}_D=\ve{L}_D(\matrix{P}_D; C_{lm}, S_{lm}; \ve{x}_{AD}). \label{L-D} 
\end{eqnarray}
\noindent
Here $\ve{L}_D$ is the torque and $\ve{S}_D$ the angular momentum in
the dynamical non-rotating frame and $\matrix{P}_D$ is a time-dependent
orthogonal matrix transforming the coordinates $\ve{X}_D$ to a terrestrial 
reference system $\ve{Y}$, where the gravitational field of the Earth is constant:
\begin{equation}
\label{xD2Y}
 Y^a=P_D^{\,ab}(T)\,X_D^b.
\end{equation}
${\cal A,B,C}$
are the principle moments of inertia of the Earth, $C_{lm}, S_{lm}$ are the
coefficients of the gravitational field of the Earth in $\ve{Y}$ and
$\ve{x}_{AD}$ are the BCRS coordinates $\ve{x}_A$ of body A (Sun,
Moon, etc.) rotated by the geodetic precession:
\begin{equation}
\label{rotation-of-ephemeris}
x^i_{AD}=R^{\,ij}(T)\,x^j_A.
\end{equation}
\noindent 
The matrix $\matrix{P}_D$ can be parametrized by Euler angles $\varphi$, $\psi$ and
$\omega$ in the usual way \citep{SMART1}.
Thus, these three angles as functions of time $T$ represent a solution of Eqs. 
(\ref{eqm-D})--(\ref{L-D}).
Note that the only difference between a purely Newtonian
solution of Earth rotation and Eqs. (\ref{eqm-D})--(\ref{L-D}) is
that the torque should be computed by using rotated positions
$\ve{x}_{AD}$ of external bodies and not the normal BCRS positions
$\ve{x}_A$. This reflects the fact that the coordinates $\ve{X}_D$ rotate
with respect to the BCRS.

The corresponding equations in the kinematically non-rotating GCRS
take the form
\begin{eqnarray}
&&{d\over dT}\,\ve{S}_K=\ve{L}_K+\bm{\Omega}_{\rm GP}\times\ve{S}_K,
\label{eqm-K}
\\
&&\ve{S}_K=\ve{S}_K(\matrix{P}_K; {\cal A,B,C}; \bm{\Omega}_{\rm GP}),
\label{S-K}
\\
&&\ve{L}_K=\ve{L}_K(\matrix{P}_K; C_{lm}, S_{lm}; \ve{x}_A).
\label{L-K}
\end{eqnarray}
\noindent
The torque $\ve{L}_K$ is defined by the same functional form as
$\ve{L}_D$, but the coordinates of external bodies are taken directly
in the BCRS. Eq. (\ref{eqm-K}) contains an additional Coriolis torque
proportional to $\bm{\Omega}_{\rm GP}$. Besides this, the angular momentum
$\ve{S}_K$ in the GCRS explicitly depends on the geodetic precession
$\bm{\Omega}_{\rm GP}$.  For details of these equations see 
\citet{klioner_etal2010} and references therein. 

From Eqs. (\ref{xK2xD}) 
and (\ref{xD2Y}) it is clear, that the solutions of these two sets of 
equations are related by
\begin{equation}
\label{PK=PD*R}
P^{\,ab}_K=P^{\,ac}_D\,R^{\,cb}.
\end{equation}

\section{\label{sec:4}Computing the geodetic precession}

To determine the effect of geodetic precession one has to compute
the matrix $\matrix{R}$. This can be done by a numerical integration of
Eqs. (\ref{Omega-GP})--(\ref{Omega-GP-R}). It should be remarked that
care has to be taken how to represent this matrix properly. To avoid the 
discontinuities that can arise when using the common Euler angles to describe 
an arbitrary rotation, we decided to use quaternions to represent this matrix. 
With matrix $\matrix{R}$, solution $\matrix{P}_D$ and Eq. (\ref{PK=PD*R}) we can 
calculate the differences $\delta\varphi$, $\delta\psi$ and $\delta\omega$ between
the Euler angles $\varphi$, $\psi$ and $\omega$ corresponding to matrix
$\matrix{P}_K$ and those corresponding to $\matrix{P}_D$.

\begin{figure} [ht!]
  \centering 
  \includegraphics[width=0.50\textwidth] {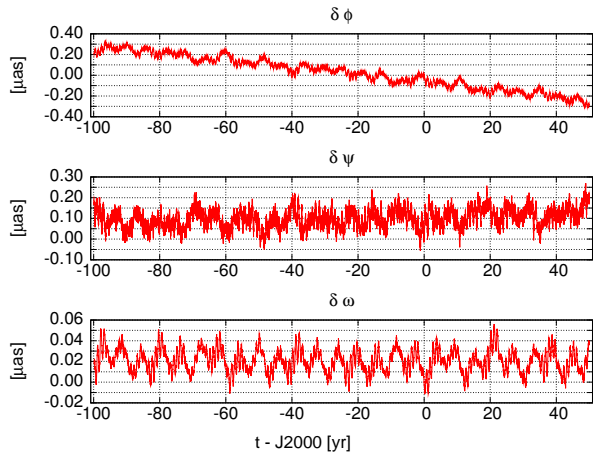} \hfill
  \caption{\label{fig:1}Differences (in \muas) for the analytical
    solution for geodetic precession derived by \citet{SMART2} and our
    numerical solution. The sign error in the SMART solution \citep{SMART2} 
    for angle $\varphi$ is corrected here.}
\end{figure}

In the process of computing and verifying the results we have found
and corrected a sign error for the correction induced by the geodetic
precession for angle $\varphi$ in \citet{SMART2}. Taking this sign
error into account the differences between the analytical solution for
$\delta\varphi$, $\delta\psi$ and $\delta\omega$ derived by
\citet{brumberg_etal1991} and \citet{SMART2} and our numerical
solution are below 1 \muas. They are shown in Fig.~\ref{fig:1}.
The remaining differences are explained by the limited accuracy of the
analytical treatment of this effect by the other authors compared to
our numerical result.

\section{\label{sec:5}Computing the GCRS solution}

According to the equations given in Section \ref{sec:3} there are two
ways to compute the matrix $\matrix{P}_K$ corresponding to the solution of the
rotational motion of the Earth with respect to the kinematically
non-rotating GCRS:

\smallskip

1. One can numerically integrate Eqs. (\ref{eqm-D})--(\ref{L-D}) and
obtain the solution $\matrix{P}_D$ with respect to dynamically non-rotating
coordinates $\ve{X}_D$. Then one can correct for geodetic precession
using the matrix $\matrix{R}$ and Eq. (\ref{PK=PD*R}) to rotate the solution into
the GCRS.

\smallskip

2. One can integrate Eqs. (\ref{eqm-K})--(\ref{L-K}) and directly obtain $\matrix{P}_K$. 

\smallskip

\noindent
Obviously, the initial conditions for Eqs. (\ref{eqm-D})--(\ref{L-D}) and (\ref{eqm-K})--(\ref{L-K}) are
again related by (\ref{PK=PD*R}) taken at the initial epoch.

We have implemented both of the above mentioned possibilities to compute $\matrix{P}_K$ and verified
that the differences in $\varphi$, $\psi$ and $\omega$ computed in the
two ways represent only numerical noise at the level of 0.001 \muas\
and less after 100 years of integration.

The implementation is done in an efficient way. The
numerical integration of the matrix $\matrix{R}$ runs for example simultaneously
with the numerical integration for $\matrix{P}_D$. The relative running times
between a purely Newtonian integration and both ways described above
are given in Table \ref{tab:cputimes}. Further details on our
numerical code and its capabilities can be found for example in
\citet{klioner_etal2008}.

\begin{table} [ht!]
 \centering
 \begin{minipage}{0.5\textwidth}
  \caption{\label{tab:cputimes}Relative CPU times for various integrations.}
  \begin{tabular}{lc}
\midrule
  Newtonian case   &   1.00 \\
  Interation of $\matrix{P}_D$ with rotated ephemeris   &  1.21 \\
  Direct integration of $\matrix{P}_K$     &  1.08
\end{tabular}
\end{minipage}
\end{table}

\begin{figure} [ht!]
  \centering 
  \includegraphics[width=0.50\textwidth] {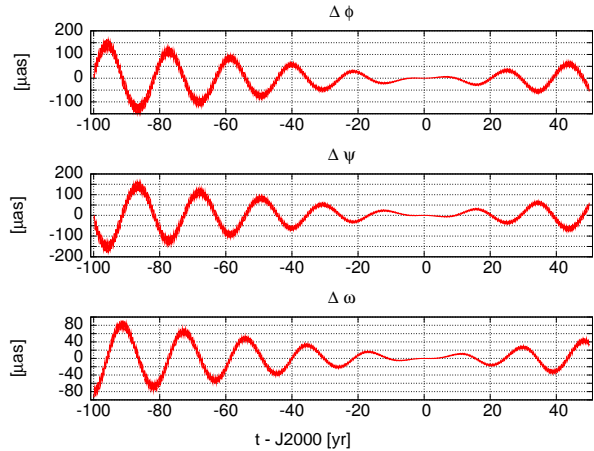} \hfill
  \caption{\label{fig:2}Differences (in \muas) for the Euler angles
    between a purely Newtonian solution and the correct solution in
    dynamically non-rotating coordinates. The latter is obtained by
    using ephemeris data rotated according to
    Eq. (\ref{rotation-of-ephemeris}).}
\end{figure}

A purely Newtonian model differs from Eqs. (\ref{eqm-D})--(\ref{L-D})
only by the positions of the solar system bodies used to compute the
torque on the Earth: the Newtonian model uses the BCRS ephemeris
directly, while for Eqs. (\ref{eqm-D})--(\ref{L-D}) one has to rotate
this ephemeris according to
Eq. (\ref{rotation-of-ephemeris}). \emph{It is this rotation that has
  never been considered before in any theory of Earth rotation, which
  represents the main source of inconsistency in the standard way of
  taking the geodetic precession into account}. The effect of the
rotation of the ephemeris on the Euler angles $\varphi$, $\psi$ and
$\omega$ is shown in Fig.~\ref{fig:2}. One finds that the error due to
this inconsistency amounts to 200 \muas\ after 100 years of
integration.

\begin{figure} [ht]
  \centering 
  \includegraphics[width=0.50\textwidth] {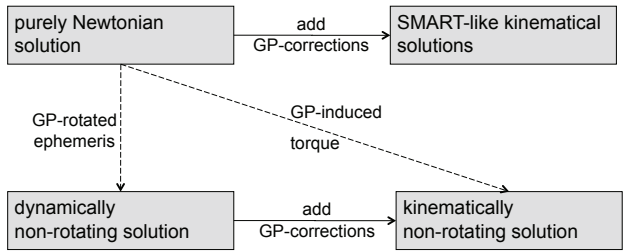} \hfill
  \caption{\label{fig:3}Schematic representation of the differences in
    the standard and the correct way to treat geodetic precession.
    ``GP'' stands for geodetic precession/nutation. Each gray block
    represents a solution. A solid arrow means: add precomputed
    geodetic precession/nutation to a solution to get a new one. A
    dashed arrow means: recompute a solution with indicated change in
    the torque model.}
\end{figure}

A summary of the interrelations between the correct solutions for the
Earth rotation in dynamically and kinematically non-rotating
coordinates as well as Newtonian and ``kinematically non-rotating''
solution derived in the standard, inconsistent way is given
schematically in Fig.~\ref{fig:3}.

\section{Difference to existing GCRS solutions}

\begin{figure} [ht!]
  \centering 
  \includegraphics[width=0.50\textwidth] {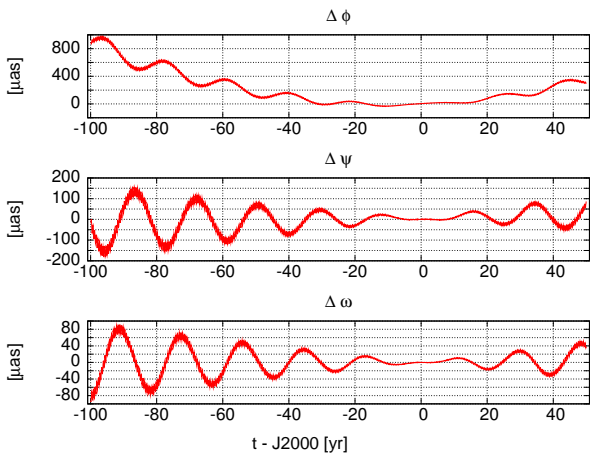} \hfill
  \caption[]{\label{fig:4}Differences (in \muas) between the published
kinematically non-rotating SMART solution and the correct
kinematically non-rotating solution derived in this study.}
\end{figure}

The difference between the Euler angles of the GCRS solution obtained
in this work and the published kinematically non-rotating SMART
solution \citep{SMART2} is given in Fig.~\ref{fig:4}.  Analysing the sources of these
differences, one can identify three components:

\begin{itemize}
\item[-] influence of the rotation of the ephemeris shown in
  Fig.~\ref{fig:2} due to the incorrect treatment of the geodetic
  precession;
\item[-] sign error in the correction for geodetic precession in $\varphi$;
\item[-] errors of the analytical SMART solution compared to the more
  accurate numerical integration, as already discussed in
  \citet{SMART2}.
\end{itemize}
It should be remarked that the above-mentioned in\-consistency is not only
restricted to the SMART solution, which we used in this study for
comparison, but is also contained in the IAU 2000A Precession-Nutation
model as described in section 5.5.1 of the IERS Conventions
\citep{iers2003}. Therefore Fig.~\ref{fig:4} allows us to
conclude that the existing GCRS solutions for rigid Earth rotation are
wrong by 1000 \muas\ in $\varphi$, 200 \muas\ in $\psi$ and 100
\muas\ in $\omega$ within 100 years from J2000. It can be shown that
these differences cannot be eliminated 
by fitting the free parameters of our model, namely the
moments of inertia, the initial Euler angles and their time derivatives.

Let us finally note that the effects of non-rigidity in the Earth
rotation and the inaccuracies of the corresponding models, e.~g. for
the atmosphere and oceans, are significantly larger than the effects
discussed in this work. Nevertheless to avoid a wrong geophysical
interpretation of the observed Earth orientation parameters, the
treatment of geodetic precession should be done along the lines
presented in this paper.

\bibliographystyle{astron}
\bibliography{bibo}

\begin{thebibliography}{}

\bibitem[\protect\astroncite{{Bretagnon} et~al.}{1998}]{SMART2}
{Bretagnon}, P., {Francou}, G., {Rocher}, P., and {Simon}, J.~L. (1998):
\newblock {\em {SMART97: a new solution for the rotation of the rigid Earth}},
\newblock Astron. Astrophys. {\bf 329}, 329-338.

\bibitem[\protect\astroncite{{Bretagnon} et~al.}{1997}]{SMART1}
{Bretagnon}, P., {Rocher}, P., and {Simon}, J.~L. (1997):
\newblock {\em {Theory of the rotation of the rigid Earth.}},
\newblock Astron. Astrophys. {\bf 319}, 305-317.

\bibitem[\protect\astroncite{{Brumberg} et~al.}{1991}]{brumberg_etal1991}
{Brumberg}, V.~A., {Bretagnon}, P., and {Francou}, G. (1991):
\newblock {\em {Analytical algorithms of relativistic reduction of astronomical
  observations.}}
\newblock in {\em Journ{\'e}es 1991: Syst{\`e}mes de R{\'e}f{\'e}rence
  Spatio-temporels}, pp 141--148.

\bibitem[\protect\astroncite{{de Sitter}}{1916}]{deSitter}
{de Sitter}, W. (1916):
\newblock {\em {Einstein's theory of gravitation and its astronomical
  consequences}},
\newblock Mon. Not. R. Astron. Soc. {\bf 76}, 699-728.

\bibitem[\protect\astroncite{{Klioner} et~al.}{2010}]{klioner_etal2010}
{Klioner}, S.~A., {Gerlach}, E., and {Soffel}, M.~H. (2010):
\newblock {\em Relativistic aspects of rotational motion of celestial bodies}
\newblock in S.~A. Klioner, P.~K. Seidelmann, and M.~H. Soffel (eds.), {\em IAU
  Symposium 261}, pp 112--123.

\bibitem[\protect\astroncite{{Klioner} et~al.}{2008}]{klioner_etal2008}
{Klioner}, S.~A., {Soffel}, M.~H., and {Le Poncin-Lafitte}, C. (2008):
\newblock {\em {Towards the relativistic theory of precession and nutation}}
\newblock in {\em Journ{\'e}es 2007: Syst{\`e}mes de R{\'e}f{\'e}rence
  Spatio-temporels}, pp 139--142.

\bibitem[\protect\astroncite{{Klioner} et~al.}{2001}]{klioner_etal2001}
{Klioner}, S.~A., {Soffel}, M.~H., {Xu}, C., and {Wu}, X. (2001):
\newblock {\em {Earth's rotation in the framework of general relativity: rigid
  multipole moments}}
\newblock in {\em The Celestial Reference Frame for the Future (Proc. of
  Journ\'ees 2007), N. Capitaine (ed.), Paris Observatory, Paris}, pp 232--238.

\bibitem[\protect\astroncite{{McCarthy} and {Petit}}{2004}]{iers2003}
{McCarthy}, D.~D. and {Petit}, G. (2004):
\newblock {\em {IERS Conventions (2003)}}, IERS Technical Note No.32, BKG,
  Frankfurt.

\bibitem[\protect\astroncite{Seidelmann}{1982}]{IAU1980}
Seidelmann, P.~K. (1982):
\newblock {\em {1980 IAU Nutation: The Final Report of the IAU Working Group on
  Nutation}},
\newblock Celest. Mech. Dyn. Astron. {\bf 27}, 79-106.

\bibitem[\protect\astroncite{{Soffel} et~al.}{2003}]{soffel_etal2003}
{Soffel}, M.~H., {Klioner}, S.~A., {Petit}, G., {Wolf}, P., {Kopeikin}, S.~M.,
  {Bretagnon}, P., {Brumberg}, V.~A., {Capitaine}, N., {Damour}, T.,
  {Fukushima}, T., {Guinot}, B., {Huang}, T., {Lindegren}, L., {Ma}, C.,
  {Nordtvedt}, K., {Ries}, J.~C., {Seidelmann}, P.~K., {Vokrouhlick{\'y}}, D.,
  {Will}, C.~M., and {Xu}, C. (2003):
\newblock {\em {The IAU 2000 Resolutions for Astrometry, Celestial Mechanics,
  and Metrology in the Relativistic Framework: Explanatory Supplement}},
\newblock Astron. J. {\bf 126}, 2687-2706.

\end{thebibliography}

\end{document}